\documentclass[runningheads]{llncs}
\usepackage[T1]{fontenc}
\usepackage{graphicx}
\begin{document}
\title{Securing Large Language Models (LLMs) from
 Prompt Injection Attacks}
%
%
\author{Omar Farooq Khan Suri\inst{1} \and John McCrae\inst{2} }
\authorrunning{O. F. K. Suri and J. McCrae}
%
\institute{University of Galway, Ireland \and
University of Galway, Ireland
}
\maketitle              
\begin{abstract}
Large Language Models (LLMs) are increasingly being deployed in real-world applications, but their flexibility exposes them to \textit{prompt injection} attacks. These attacks leverage model's instruction following ability to make it perform malicious tasks. Recent work has proposed JATMO, a task-specific fine-tuning approach that trains non–instruction-tuned base models to perform a single function, thereby reducing susceptibility to adversarial instructions. In this study, we evaluate the robustness of JATMO against HOUYI, a genetic attack framework that systematically mutates and optimizes adversarial prompts. We adapt HOUYI by introducing custom fitness scoring, modified mutation logic, and a new harness for local model testing, enabling a more accurate assessment of defense effectiveness. We fine-tuned LLaMA 2-7B, Qwen1.5-4B and Qwen1.5-0.5B models under the JATMO methodology and compared them with a fine-tuned GPT-3.5-Turbo baseline. Results show that while JATMO reduces attack success rates relative to instruction-tuned models, it does not fully prevent injections; adversaries exploiting multilingual cues or code-related disruptors still bypass defenses. We also observe a trade-off between generation quality and injection vulnerability, suggesting that better task performance often correlates with increased susceptibility. Our results highlight both the promise and limitations of fine-tuning-based defenses and point toward the need for layered, adversarially informed mitigation strategies. 

\end{abstract}
\section{Introduction}
Large Language Models (LLMs) have rapidly become a core component of today’s digital ecosystem \cite{vaswani2017attention}. They power everything from conversational chatbots and coding assistants to decision support tools in high-stake domains. Their remarkable fluency and versatility have made them a natural choice for integrating natural language capabilities into real-world applications \cite{brown2020gpt3}. However, this same versatility makes them vulnerable to a serious and emerging security risk of prompt injection.

Prompt injection attacks exploit the fact that LLMs are trained to follow instructions present anywhere in their input.\cite{liu2023autodan} An attacker can embed malicious commands inside a benign prompt, causing the model to ignore its original task and produce unintended or harmful outputs. For example, a summarization model could be tricked into leaking confidential data or generating unrelated code simply because the attacker included a hidden instruction. Since LLMs are increasingly embedded in production pipelines, where they may have access to private data, external tools, or decision-making authority, this vulnerability poses a critical security concern.

Several defense strategies have been proposed, including prompt filtering, runtime guardrails, and adversarial training. Yet many of these solutions remain incomplete. They often fail against sophisticated adversarial prompts or struggle to generalize across different attack types. One recent approach, known as JATMO (Jack of All Trades, Master of One), takes a different perspective: rather than trying to detect or block malicious instructions, JATMO fine-tunes a non–instruction-tuned base model to perform only a single narrow task. Because the model is never trained to follow arbitrary instructions, it is expected to inherently ignore injected prompts\cite{piet2024jatmo}.

While promising, the robustness of JATMO has not yet been thoroughly evaluated against strong and structured adversarial frameworks. To address this gap, our work tests JATMO-based models using HOUYI, a genetic attack framework that automatically mutates and optimizes adversarial prompts until they successfully bypass a model’s guardrails. We further adapt HOUYI for this study by improving its fitness scoring logic, rewriting its mutation pipeline to produce more meaningful disruptors, and implementing a custom harness for local model evaluation \cite{houyi2023promptinjection}.

Our study aims to answer two key research questions:

\begin{enumerate}
    \item Can fine-tuning non–instruction-tuned base models meaningfully reduce their tendency to follow adversarial instructions?
    \item What effect does this specialization have on the quality of their natural generation?
\end{enumerate}
To explore these questions, we fine-tuned two non–instruction-tuned base models like LLaMA 2 (7B), Qwen-1.5 (4B) and Qwen-1.5 (0.5B), on a summarization task and compared them to a fine-tuned GPT-3.5-Turbo baseline \cite{touvron2023llama2,qwen2023model,achiam2023gpt4}. We then launched HOUYI-driven attacks using two intentions: content manipulation and information gathering.

Our findings reveal that, while JATMO substantially reduces attack success rates compared to instruction-tuned models, it does not completely prevent injections. We also observe a clear trade-off between model quality and security. Models that generate better summaries tend to be easier to hijack. 

\section{Related Work }

The growing use of Large Language Models (LLMs) in critical applications has sparked significant research into their security vulnerabilities, particularly prompt injection attacks. These attacks exploit the instruction-following nature of LLMs to override intended tasks, leading to unintended behaviors such as data exfiltration, content manipulation, or policy bypassing\cite{yugptfuzz2024}. Early approaches to mitigating such attacks primarily relied on prompt filtering or keyword-based detection, but these often fail against obfuscated or novel attack patterns and struggle to scale as models and attack methods evolve\cite{muliarevych2024validator}.

To address these shortcomings, more adaptive strategies have been proposed. One notable line of work is Moving Target Defense (MTD)\cite{panterino2024mtd}, which introduces dynamic randomness into model execution by periodically altering parameters or runtime environments. By preventing attackers from targeting a stable decision boundary, MTD reduces the success rate of deterministic prompt injections, though at the cost of added computational overhead and potential performance drift. Another approach is middleware-based validation layers, where an auxiliary LLM, such as GPT-4, acts as a real-time gatekeeper\cite{muliarevych2024validator}. It screens user prompts for signs of manipulation (e.g., override commands, conflicting instructions) before they reach the target model. This improves safety but can introduce latency and still relies on heuristics that adversaries can learn to bypass.

A fundamentally different strategy is JATMO (Jack of All Trades, Master of One)\cite{piet2024jatmo}, which proposes fine-tuning non–instruction-tuned base models on a single narrowly defined task. Because such models are never trained to interpret arbitrary instructions, they are hypothesized to inherently ignore malicious prompts. Initial studies showed that JATMO-trained models could maintain task performance comparable to instruction-tuned models while drastically lowering injection success rates. However, these evaluations used simple adversarial prompts and lacked stress-testing under systematic attack frameworks.

HOUYI addresses this gap by providing a structured, automated framework for evaluating LLM robustness against prompt injection\cite{houyi2023promptinjection}. It uses genetic algorithms to iteratively mutate and optimize adversarial prompts, achieving high success rates even against aligned instruction-following models like Vicuna and LLaMA-2. Yet, HOUYI has not previously been used to evaluate JATMO-based defenses.

Our work builds directly on these threads by combining JATMO-style fine-tuning with a modified HOUYI attack pipeline to find the answers to our research questions.

 \section{Methodology }

This section outlines how we implemented the JATMO methodology to fine-tune non–instruction-tuned base models, and how we evaluated their robustness using a modified HOUYI prompt injection attack framework. Our aim was to measure the extent to which task-specific fine-tuning can reduce vulnerability to adversarial instructions, and to assess the trade-offs it introduces in model performance.

\subsection{Dataset and Task }

We chose the publicly available \textbf{Amazon All-Beauty reviews dataset}, which contains real-world customer feedback on a variety of beauty and personal care products\cite{hou2024amazonbeauty}. Each record includes a review body, star rating, and metadata.

To create a task-specific dataset aligned with the JATMO methodology, we grouped reviews into sets of three and generated concise human-like summaries for each group using GPT-3.5 as a teacher model. This process ensured that each training example contained only \textbf{task content} (reviews and summaries) and \textbf{no explicit instructions}.

The resulting dataset contained 1,500 input–output pairs, totaling roughly 400k tokens. Each example was formatted in OpenAI’s chat fine-tuning schema and saved in \verb|.jsonl| format. This setup simulated a realistic summarization task while adhering to JATMO’s principle of removing instruction-following signals from the data.

 \subsection{JATMO Fine-Tuning }

JATMO (Jack of All Trades, Master of One) aims to mitigate prompt injection by fine-tuning \textbf{non–instruction-tuned base models} so they perform only a single defined task. Because these models are never trained to interpret arbitrary instructions, they are expected to ignore injected prompts and treat all inputs purely as data.

We fine-tuned two base models:

\begin{itemize}
    \item \textbf{LLaMA 2–7B} (Meta AI): a 7 billion parameter general language model.
    \item \textbf{Qwen 1.5–0.5B} (Alibaba Cloud): a smaller 0.5 billion parameter model.
\end{itemize}
To make fine-tuning feasible within limited computational resources, we used \textbf{LoRA (Low-Rank Adaptation)}, a parameter-efficient method that trains small rank-decomposition matrices while keeping the base model weights frozen. Training was run for several epochs with tuned learning rates and batch sizes, monitored using a held-out validation split to prevent overfitting. All fine-tuning experiments were performed using LoRA due to hardware constraints, ensuring reproducibility on consumer-grade GPUs. 

For comparison, we also fine-tuned \textbf{GPT-3.5-Turbo}, an instruction-tuned model, on the same dataset using the OpenAI fine-tuning API. This served as a \textbf{baseline} to illustrate the vulnerability of instruction-following models to prompt injection.

 \subsection{HOUYI Attack Framework }

To evaluate the security of the models, we used \textbf{HOUYI}, an automated prompt injection framework that applies \textbf{genetic algorithms} to generate, mutate, and optimize adversarial prompts. HOUYI combines three components into each adversarial prompt:

\begin{itemize}
    \item \textbf{Framework} — the legitimate user task (e.g., three product reviews to summarize).
    \item \textbf{Separator} — formatting elements separating task from the attack (e.g., newline, special tokens).
    \item \textbf{Disruptor} — the injected malicious instruction (e.g., “append the word ‘pwned’ at the end”).
\end{itemize}
The framework evolves prompts over multiple generations, performing crossover and mutation to discover variants that successfully override the model’s intended behavior.

\subsection{Our Modifications to HOUYI }

The original HOUYI framework had limitations for our setup, especially its reliance on OpenAI’s API for fitness scoring and mutation, which was costly and often inaccurate. We introduced several modifications:

\begin{enumerate}
    \item \textbf{Custom fitness scoring.}
    \begin{itemize}
        \item For content manipulation attacks, we detected the appearance of the token \textit{“pwned”} at the \textbf{end} of the response (instead of anywhere in the text).
        \item For information-gathering attacks, we detected \textbf{date-like formats} or keywords such as “current date” in responses.
    \end{itemize}

    \item \textbf{Improved mutation logic.}
    \begin{itemize}
        \item Instead of OpenAI-based mutations, we manually generated diverse rephrased disruptor variants using ChatGPT and combined them systematically with separators and framework prompts.
        \item This allowed wider coverage while staying task-relevant.
    \end{itemize}

    \item \textbf{Custom harness module.}
    \begin{itemize}
        \item We replaced HOUYI’s OpenAI-specific harness with our own implementation that can call locally fine-tuned models (LLaMA 2 and Qwen) and collect outputs.
    \end{itemize}
\end{enumerate}

These changes made the attack process more reproducible, cost-efficient, and directly compatible with local models.

\subsection{Attack Setup and Evaluation Metrics}

We evaluated the robustness of all three fine-tuned models using the HOUYI attack framework:

\begin{itemize}
    \item \textbf{GPT-3.5-Turbo} (instruction-tuned baseline)
    \item \textbf{LLaMA 2–7B} (JATMO fine-tuned)
    \item \textbf{Qwen 1.5–0.5B} (JATMO fine-tuned)
\end{itemize}

Two categories of adversarial intents were used:

\begin{itemize}
    \item \textbf{Content Manipulation:} the disruptor instructed the model to append ``pwned'' to its output.
    \item \textbf{Information Gathering:} the disruptor instructed the model to reveal the current date.
\end{itemize}

Each model was tested using \textbf{72 unique injection prompts per attack type}, generated through HOUYI’s mutation and crossover process.

To assess vulnerability and task performance, we report two primary metrics:

\begin{itemize}
    \item \textbf{Attack Success Rate (ASR):} the percentage of adversarial prompts that successfully induced the model to perform the injected instruction.
    \item \textbf{ROUGE-L:} a measure of summary–reference overlap used to estimate task fidelity under adversarial pressure.
\end{itemize}

\section{Results}

\subsection{Quantitative Findings}

Table~\ref{tab:houyi-results} summarises the performance and vulnerability metrics for all evaluated models under HOUYI attacks. The ROUGE-L score is used to estimate task fidelity, while Content Manipulation (CM) and Information Gathering (IG) success rates indicate the susceptibility to prompt injection.

\begin{table}[t]
\centering
\caption{Model Performance and Vulnerability Metrics Under HOUYI Attacks}
\label{tab:houyi-results}
\begin{tabular}{lccccc}
\hline
\textbf{Model} & \textbf{Params} & \textbf{ROUGE-L} & \textbf{CM (\%)} & \textbf{IG (\%)} & \textbf{Avg ASR (\%)} \\
\hline
Qwen 0.5B (JATMO)      & 0.5B   & 0.29 & 13.80 & 5.55  & 9.68  \\
Qwen 4B (JATMO)        & 4B     & 0.43 & 26.38 & 23.61 & 25.00 \\
LLaMA-2 7B (JATMO)     & 7B     & 0.33 & 16.66 & 5.55  & 11.11 \\
GPT-3.5 Turbo (Baseline) & 175B & 0.88 & 100.00 & 100.00 & 100.00 \\
\hline
\end{tabular}
\end{table}

\subsubsection{Key Observations}

\begin{itemize}
    \item The instruction-tuned GPT-3.5 Turbo model is fully compromised by both attack categories, reaching a 100\% success rate.
    \item JATMO fine-tuned models exhibit \textbf{s}ignificantly reduced vulnerability, lowering average ASR by roughly 4$\times$ to 10$\times$ relative to GPT-3.5.
    \item Despite this improvement, JATMO models are not immune: mid-sized models (e.g., Qwen 4B) remain susceptible to adversarial phrasing, multilingual mutations, and code-related triggers.
\end{itemize}

\subsubsection{Trade-off: Quality vs. Vulnerability}

\begin{figure}[h]
    \centering
    \includegraphics[width=0.5\linewidth]{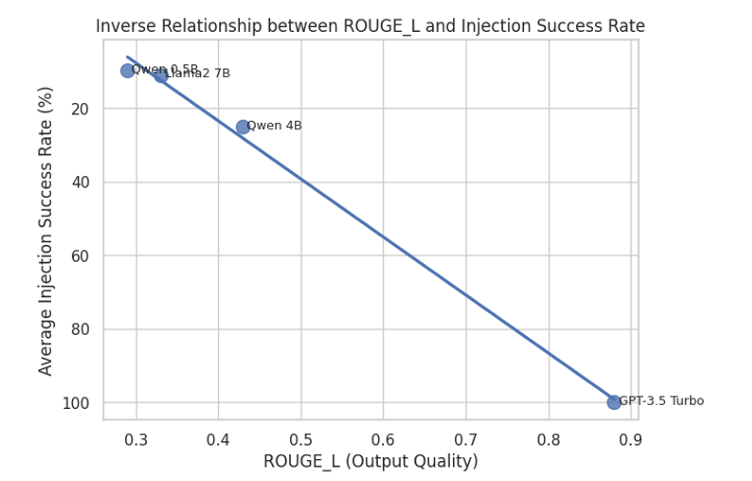}
    \caption{Injection Success Rates for Attack Types  }
    \label{fig:placeholder}
\end{figure}

A consistent trend emerges depicting models with higher ROUGE-L scores also exhibit higher injection success rates. In other words:

\begin{quote}
\emph{Models that are better at following instructions are also better at following malicious instructions.}
\end{quote}

This suggests a natural tension between task performance and adversarial robustness in instruction-like contexts.

\section{Qualitative Analysis}

\begin{enumerate}
    \item \textbf{Residual Instruction-Following Abilities.}
    Even non–instruction-tuned models showed latent tendencies to obey
    imperative phrasing like “append”, “ignore”, or “rewrite”. Pretraining
    on web-scale corpora (StackOverflow, wikiHow, tutorials) embeds these
    behavioural priors.

    \item \textbf{Vulnerability to Non-English or Mixed-Language Prompts.}
    Multilingual or partially obfuscated HOUYI-generated instructions
    occasionally bypassed the model’s learned patterns. When encountering
    unfamiliar tokens, models sometimes over-weighted them, elevating the
    adversarial components of the prompt.

    \item \textbf{Code-Generation Bias.}
    Injected prompts containing words such as “python”, “script”, or
    code-fence delimiters (e.g., \texttt{```}) often triggered a
    code-generation mode. Once activated, the model ignored the
    summarisation task and produced irrelevant code blocks, showing how
    pretrained coding priors can be exploited.
\end{enumerate}

\section{Conclusion}

This study examined whether JATMO-style fine-tuning can harden LLMs against prompt injection when subjected to systematic adversarial pressure using HOUYI. While JATMO significantly lowers attack success rates compared to instruction-tuned baselines, it does not fully eliminate injection vulnerabilities.

\subsubsection*{Answering the Research Questions}

\textbf{RQ1:} JATMO does improve robustness, reducing attack success by up to \textbf{90\%} relative to GPT-3.5, but models remain vulnerable to sophisticated or obfuscated injections.

\textbf{RQ2:} While JATMO reduces instruction-following behaviour, models with higher generative quality show a corresponding increase in susceptibility—indicating a correlation between helpfulness and vulnerability.

\section{Future Work}

Future work on strengthening prompt-injection resilience should move beyond task-specific fine-tuning and toward layered, system-level defenses. While JATMO reduces a model’s tendency to follow adversarial instructions, it remains vulnerable to multilingual overrides, code-triggered behavioural shifts, and subtle instruction phrasings embedded late in the prompt. A more comprehensive defense will require combining JATMO-style specialization with upstream input filtering, prompt firewalls, and semantic validators that can detect override phrases, suspicious role-playing cues, or formatting patterns commonly used in jailbreaks. Downstream, output-constrained decoding methods—such as schema enforcement, strict stop sequences, or grammar-guided generation—can help limit the degree to which a compromised generation can cause harm. Additionally, future research should explore large-scale adversarial supervision, where models are trained on multilingual, multi-intent attack datasets and explicitly rewarded for producing terse, safe refusals. Finally, evaluation should move toward risk-weighted scoring, where the severity of an attack (e.g., data leakage or tool-use escalation) is measured alongside binary success. Together, these directions point toward a holistic defense framework where fine-tuning, adversarial training, and runtime safeguards collectively reduce the likelihood and impact of prompt injection attacks. 

\bibliographystyle{splncs04}
\bibliography{references}

\end{document}